\documentclass[preprint,superscriptaddress,11pt]{revtex4-1}
\usepackage{graphicx}
\usepackage{dcolumn}
\usepackage{color}
\usepackage{times}
\usepackage{bm}
\usepackage{amssymb}
\usepackage{amsmath}
\usepackage{epsfig}
\usepackage{epstopdf}
\begin{document}
\renewcommand{\thefootnote}{\fnsymbol {footnote}}

\title{Entropic uncertainty relations for Markovian and non-Markovian processes under a structured bosonic reservoir}

\author{Dong Wang} \email{dwang@ahu.edu.cn (D.W.)}
\affiliation{School of Physics \& Material Science, Anhui University, Hefei
230601, China}
\affiliation{National Laboratory for Infrared Physics, Shanghai Institute of Technical Physics, Chinese Academy of Sciences, Shanghai 200083, China}

\author{Ai-Jun Huang}
\affiliation{School of Physics \& Material Science, Anhui University, Hefei
230601, China}

\author{Ross D. Hoehn}
\affiliation{Department of Chemistry, Department of Physics
and Birck Nanotechnology Center, Purdue University, West Lafayette, IN 47907 USA}
\affiliation{Qatar Environment and Energy Research Institute(QEERI), HBKU, Qatar Foundation, Doha, Qatar}

\author{Fei Ming}
\affiliation{School of Physics \& Material Science, Anhui University, Hefei
230601, China}

\author{Wen-Yang Sun} \author{Jia-Dong Shi}
\author{Liu Ye}\email{yeliu@ahu.edu.cn (L.Y.)}
\affiliation{School of Physics \& Material Science, Anhui University, Hefei
230601, China}

\author{Sabre Kais}

\affiliation{Department of Chemistry, Department of Physics
and Birck Nanotechnology Center, Purdue University, West Lafayette, IN 47907 USA}
\affiliation{Qatar Environment and Energy Research Institute(QEERI), HBKU, Qatar Foundation, Doha, Qatar}

\date{\today}

\begin{abstract}{\bf
The uncertainty relation is a fundamental limit in quantum mechanics and is of great importance to quantum information processing as it relates to quantum precision measurement. Due to interactions with the surrounding environment, a quantum system will unavoidably suffer from decoherence. Here, we investigate the dynamic behaviors of the entropic uncertainty relation of an atom-cavity interacting system under a bosonic reservoir during the crossover between Markovian and non-Markovian regimes. Specifically, we explore the dynamic behavior of the entropic uncertainty relation for a pair of incompatible observables under the reservoir-induced atomic decay effect both with and without quantum memory. We find that the uncertainty dramatically depends on both the atom-cavity and the cavity-reservoir interactions, as well as the correlation time, $\tau$, of the structured reservoir.  Furthermore, we verify that the uncertainty is anti-correlated with the purity of the state of the observed qubit-system. We also propose a remarkably simple and efficient way to reduce the uncertainty by utilizing quantum weak measurement reversal. Therefore our work offers a new insight into the uncertainty dynamics for multi-component measurements within an open system, and is thus important for quantum precision measurements.}
\end{abstract}

\maketitle

The uncertainty principle, originally proposed by Heisenberg \cite{Heisenberg}, is a fascinating aspect of quantum mechanics. It sets a bound to the precision for simultaneous measurements regarding a pair of incompatible observables, \emph{e.g.} position ($\hat{x}$) and momentum ($\hat{p}$). Later, the uncertainty principle was generalized, by Kennard \cite{E.H. Kennard} and Robertson \cite{H. P. Robertson} as applying to an arbitrary pair of
non-commuting observables (say ${\hat{\cal P}}$ and ${\hat{\cal Q}}$)
where the standard deviation is given as
\begin{align}
\Delta_\rho{{\hat{\cal P}}}\cdot\Delta_\rho{ \hat{\cal Q}}\geq\frac12|\langle[{ \hat{\cal P}},{\hat{\cal Q}}]\rangle|_{\rho}
\end{align}
for a given system, $\rho$, where the variance is given as $\Delta_\rho {{\cal X}}=\sqrt{{\langle{\cal X}^2\rangle_\rho-\langle{\cal X}\rangle^2_\rho}}$,
$\langle \bullet\rangle$ denotes the expectation value of the observable, and $[\hat{\cal P},\hat{Q}]={ \hat{\cal P}\hat{\cal Q}}-{\hat{\cal Q}\hat{\cal P}}$ denotes the commutator. Importantly, the standard deviation in Robertson's relation is not always an optimal measurement for the uncertainty as the right-hand side of the relation depends on the state $\rho$ of the system, which will lead to a trivial bound if the operators ${\hat{\cal P}}$ and ${\hat{\cal Q}}$ do not commute. In order to compensate for this, Deutsch \cite{D. Deutsch} put forward an alternative inequality of the form
\begin{align}
S^\rho({ \hat{\cal P}})+S^\rho({ \hat{\cal Q}})\geq 2 {\rm log}_2{\left(\frac2{1+\sqrt{c}}\right)}
\label{Eq.2}
\end{align}
for any pair of non-degenerate observables $\hat{\cal P}$ and $\hat{\cal Q}$ in terms of Shannon entropy, \emph{i.e.} the so-called entropic uncertainty relation (EUR). To be explicit, the Shannon entropy is given by $S^\rho(\hat{P})=-\sum_i p_i{\rm log}p_i$, where $p_i=\langle\psi_i|\rho|\psi_i\rangle$; the parameter $c$ in Eq. (\ref{Eq.2}) weighs the maximum value of the overlap between observables $\hat{\cal P}$ and $\hat{\cal Q}$, which can be mathematically expressed as $c={\rm max}_{ij}|\langle\psi_i|\varphi_j\rangle|^2$, with $|\psi_i\rangle$ and $|\varphi_j\rangle$ being the eigenstates of $\hat{\cal P}$ and $\hat{\cal Q}$. Obviously, yet Remarkable, that the lower bound is now independent on the state of the given system.
Later, Kraus \cite{K. Kraus}, as well as Maassen and Uffink \cite{H. Maassen}
made a significant improvement by refining Deutsch's result to
\begin{align}
S^\rho({ \hat{\cal P}})+S^\rho({ \hat{\cal Q}})\geq - {\rm log}_2 c=: B_{KMU},
\label{Eq.3}
\end{align}
where the largest uncertainty can be obtained for two arbitrary mutually unbiased observables. More recently, Coles and Piani \cite{P. J. Coles} have obtained an optimal solution with form
\begin{align}
S^\rho({ \hat{\cal P}})+S^\rho({ \hat{\cal Q}})\geq
-{\rm log}_2 c + {1-\sqrt{c} \over 2} {\rm log}_2 {c/{\tilde{c}}}
=: B_{CP},
\label{Eq.4}
\end{align}
with $\tilde{c}$ being the second largest value of $\{|\langle\psi_i|\varphi_j\rangle|^2\}$ for all values of $i$ and $j$. It is obvious that the bound $B_{CP}\geq -{\rm log}_2 c$ holds, which implies Eq. (\ref{Eq.4}) offers a tighter bound when compared with the former iterations.

In fact, the importance of the uncertainty principle is that it reflects the ability of stored quantum information within quantum memory to reduce or eliminate the uncertainty associated with a measurement on a second particle entangled to the quantum memory \cite{C. F. Li,R. Prevedel}.
Moreover, EUR has been established as a powerful tool for various applications, including: security analysis for quantum communication \cite{P. J. Coles},
entanglement witness \cite{M. Hu,M. Berta,H. M. Zou},
probing quantum correlation \cite{M. L. Hu,M. Hu2}, quantum speed limit \cite{D. Mondal,D. P. Pires}, and steering Bell's inequality \cite{J. Schneeloch}.
Additionally, there have been several expressions for the optimal outcome of EUR associated with two-component or multiple measurements \cite{Pati,F. Adabi,Xiao}.
Notably, due to interacting with a noisy environment, the quantum system will suffer from decoherence, thereby inflating the entropic uncertainty to some extent. Therefore, it is of fundamentally importance to clarify
how environmentally-induced decoherence affects the uncertainty of measurements. Till now, there have been some observations with respect to the entropic uncertainty under the influence of various types of dissipative environments \cite{Z. Y. Xu, H. M. Zou, J. Zhang, S. Liu, Y. J. Zhang, Huang,Q. Sun}.  Recently, Karpat \emph{et al.} \cite{Karpat} proposed an interesting argument that the memory effects can straightforward manipulate EUR's lower bound
in a practical scenario.

It is well known that, any environment can be classified as either Markovian (information stored in the qubit system flows one-way from the system to the environment) or non-Markvian
(information stored in the qubit system is capable of bidirectional flow between the system and the environment). Here, we aim
to understand how a structured environment affects the EUR as it undergoes a crossover between non-Markovian and Markovian regimes. The model herein considered is a two-level atomic system coupled to a composite environment, which consists of a single cavity mode and a structured reservoir. The model is simple yet sophisticated enough for our purpose. It should be noted that non-Markovian dynamics for the qubit-cavity model has been studied theoretically \cite{T.T. Ma} and demonstrated experimentally \cite{K. H. Madsen} beyond the non-Markovian regime. For a reservoir with an Ornstein-Uhlenbeck type of correlation function, the reservoir correlation time may be described with a single parameter,
conveying the reservoir's decay time. Composite environments include several time scales denoting the information exchange between the two subsystems, as well as between the system and the environment. However, the single parameter method is not generalizable to composite environments. Therefore, we investigate a several-parameter regime for the cavity-reservoir coupling strength and show
how these parameters affect the EUR. Remarkably, we found that the dissipation of the external environment caused quantitative fluctuations in the value of the entropic uncertainty. In particular, we also provide a simple and efficient way to decrease the uncertainty by leveraging the degradation of the initial state of the subsystem induced by this hierarchical environment via quantum weak measurement reversals.

\begin{figure*}[h]
\centering
\includegraphics[width=8cm]{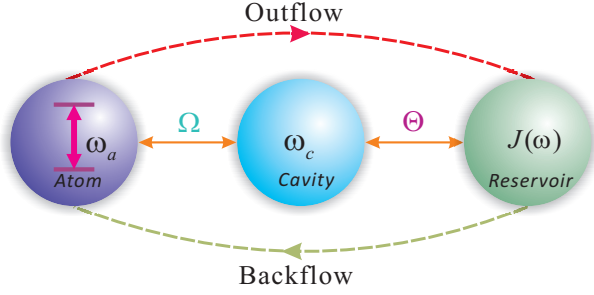}
\caption{A schematic of information flow within the composite system consisting of the atom, single-mode cavity and reservoir with Lorentz spectrum $J(\omega)$. Explicitly, the atom is coupled with the cavity by coupling constant, $\Omega$, and the cavity is coupled with a structured reservoir with an additional coupling constant, $\Theta$.}
\label{fig.1}
\end{figure*}

\bigskip

{ \large \noindent {\bf Results}}\\
\noindent { \bf Systemic dynamics.} \ \ \
Herein we consider a model system consisting of an atom (a qubit), a single-mode cavity and treat the environment as a structured bosonic reservoir. As illustrated in Fig. \ref{fig.1}, information can flow between the atom, the cavity and the reservoir. Explicitly, during a Markovian evolution the information will outflow from the qubit to environment which consists of the cavity and reservoir.  On the contrary, if the system exists within a non-Markovian regime, information will not only outflow but also backflow from the qubit to the hierarchical environment. The system can be described by the Hamiltonian
\begin{align}
&{\mathcal{H}}_{S}={\mathcal{H}}_0+{\mathcal{H}}_I,
\label{eq:hamiltonian}
\end{align}
where
\begin{align}
\mathcal{H}_0=\frac{\omega_a} 2\sigma_z+\omega_c a^{\dagger} a+\sum_{j=0}^{\infty}\omega_jb_j^{\dagger}b_j
\label{eq:freehamiltonian}
\end{align}
is the free Hamiltonian of the composite system consisting of an atom, a cavity and a structural reservoir.  Within Eqs. (\ref{eq:hamiltonian}) and (\ref{eq:freehamiltonian}), $\omega_a$, $\omega_c$ and $\omega_j$ denote the transition frequency of the atom,  the transition frequency of the cavity, and frequency of the $j$th mode of the reservoir, respectively. The Pauli operator $\sigma_z=|e\rangle\langle e|-|g\rangle\langle g|$ with $|e\rangle$ and $|g\rangle$ representing the excited  and  ground states, respectively.  $a^{\dagger}(a)$ and $b_j^{\dagger}(b_j)$ denote the creation (annihilation) operators for the cavity and the $j$th mode of the reservoir, respectively. Finally, $\mathcal{H}_I$ denotes the interaction Hamiltonian for both the atom-cavity and the cavity-reservoir.
In the interaction picture --- under the resonance condition ($\omega_a=\omega_c=\varpi$) --- the interaction Hamiltonian, $\mathcal{H}_I$, can be written as
\begin{align}
&{\cal H}_{I}=\Omega(\sigma^+a+\sigma^-a^{\dagger})
+\sum_{j=0}^{\infty}\Delta_j(ab^{\dagger}_je^{i\delta_jt}+a^{\dagger}b_je^{-i\delta_jt}).
\end{align}
Within the above, $\sigma^{+}=|e\rangle\langle g|$ and $\sigma^-=|g\rangle\langle e|$ are the upper and lower operator, respectively; $\Omega$ is the atom-cavity coupling strength, $\Delta_j$ is the coupling strength between the  cavity mode and the $j$th mode of the reservoir, and $\delta_j=\omega_j-\varpi$ describes the detuning of the cavity and the reservoir.  We assume that the reservoir has a Lorentzian spectrum $J(\omega)=\frac \Theta {2\pi} \frac  {\gamma^2}{(\varpi-\omega)^2+\gamma^2}$.
In this case, the correlation function of the reservoir is given by $\alpha(t,s)=\frac{\Theta\gamma}{2}e^{-\gamma|t-s|}$, and the correlation
(or memory time) is $\tau=\gamma^{-1}$. When $\gamma$ goes to infinity, the model environment tends to a reservoir possessing no memory effect.  Under these assumptions, we  obtain reduced dynamics for the atomic state, which is given as (see Method section for details)
\begin{align}
\rho(t)=\left(
\begin{array}{cc}
\rho_{ee}(t) & \rho_{eg}(t) \\
\rho_{eg}^{*}(t) & 1-\rho_{ee}(t)
\end{array}
\right),\label{eq.7}
\end{align}
where $\rho_{ee}(t)=\rho_{ee}(0)|\Gamma(t)|^2$ and $\rho_{eg}(t)=\rho_{eg}(0)\Gamma(t)$ with
\begin{align}
\Gamma(t)=L^{-1}[\Upsilon(p)]; \quad \Upsilon(p)=\frac{2p(p+\gamma)+{\Theta\gamma}}{2(p^2+\Omega^2)(p+\gamma)+p\Theta\gamma}.
\label{Eq.9}
\end{align}
Where $L^{-1}$ is the canonical inverse Laplace transformation.

\bigskip

\noindent {\bf EUR under a reservoir with memory.}
Assume the initial state of the atom to be an arbitrary pure state represented by $|\Psi_{\rm in}(\theta,\phi)\rangle={\rm cos}(\theta)|e\rangle+{\rm sin}(\theta) e^{i\phi}|g\rangle$, with $\theta\in[0,\pi/2]$ and $\phi\in[0,\pi]$.
A Markovian evolution can always be represented by a dynamic semigroup of completely positive and trace-preserving maps. These properties guarantee the contractiveness of the trace distance
\begin{align}
D(\rho_1(t),\rho_2(t))=\frac12{\rm Tr}|\rho_1(t)-\rho_2(t)|.
\label{Eq.10}
\end{align}
In Eq. (\ref{Eq.10}), the general form of the magnitude is $|\chi|=\sqrt{\chi^{\dag} \chi}$ between an arbitrary state $\rho_1$ and another state $\rho_2$. Note that, a Markovian process is unable to increase $D(\rho_1,\rho_2)$ at any time step.
In other words, a Markovian process either decreases or maintains the trace distance.
Essentially, the reduction of the trace distance is indicative of a reduction in the distinguishability between the two states; this could be interpreted as an outflow of information from the qubit subsystem to the environment.
Accordingly, the increase of trace distance can be understood as a backflow of information into the atomic system of interest, which is characterized by non-Markovian evolution.
Hence, the violation of the contractiveness of the trace distance would signify the on-set of
non-Markovian dynamics in the system. To be explicit, non-Markovianity \cite {Addis} in a system can be measured by
\begin{equation}
{\cal N}=\underset{\rho_1(0),\rho_2(0)}{\rm max}\int_{\sigma>0} dt \sigma(t,\rho_1(0),\rho_2(0)),
\label{Eq.11}
\end{equation}
where $\sigma(t,\rho_1(0),\rho_2(0))=\frac d{dt}D(\rho_1(t),\rho_2(t))$ is the rate of change of the trace distance as expressed by Eq. (\ref{Eq.10}).
\begin{figure*}[h]
\centering
\includegraphics[width=14cm]{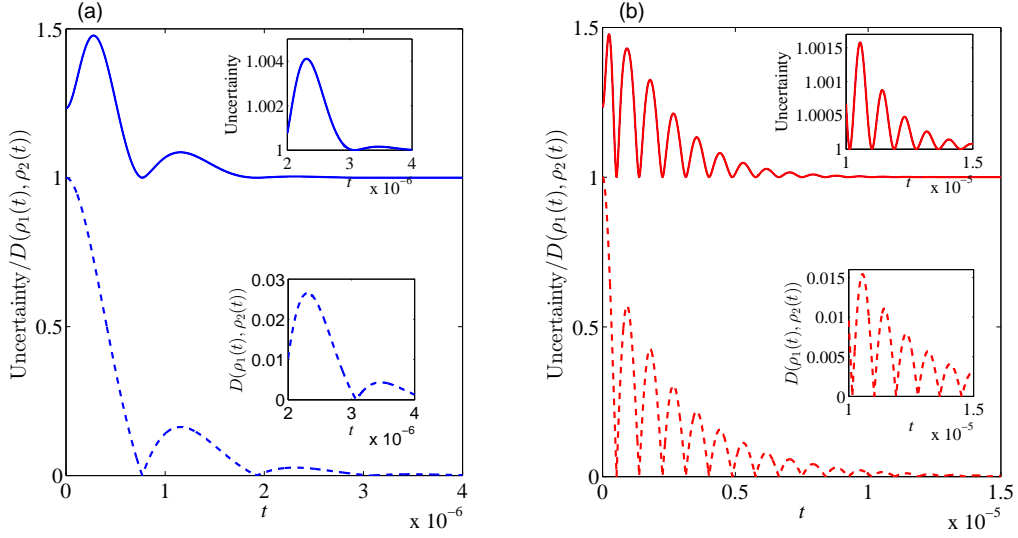}
\caption{The trace distance, $D(\rho_1,\rho_2)$, and entropic uncertainty as a function of $t$ for an initial state constructed with $\theta=\pi/4$ and $\phi=\pi/8$; we have set $\Omega =\Theta =\pi\times10^{6}$ Hz. In Graphs (a) and (b), $ \gamma= 1000\Omega$ and $\gamma= \Omega$, respectively. Within each graph, the solid line represents the uncertainty while the dashed line represents the trace distance in the two graphs.}
\label{fig.2}
\end{figure*}

To clearly display the evolution of an atomic system under the reservoir with memory, we may utilize an optimal pair of states --- ($\rho_1(0)=|+\rangle\langle+|,$ $\rho_2(0)=|-\rangle\langle-|$) --- as the two initial states, where $|\pm\rangle=(|e\rangle\pm|g\rangle)/\sqrt2$ as verified by previous works \cite{Z. He,Zhong-Xiao Man}. Thereby, after some calculations, the trace distance can be derived as:
\begin{align}
D(\rho_1(t),\rho_2(t))=|\Gamma(t)|,
\end{align}
where $\Gamma(t)$ is taken as Eq. (\ref{Eq.9}) and satisfies $-1\leq \Gamma(t)\leq1$.
Incidentally, henceforth an abbreviation (TD) shall be used to represent the trace distance, $D(\rho_1(t),\rho_2(t))$, calculated under the two optimal initial states $\{|+\rangle\langle+|,|-\rangle\langle-|\}$.
In this case, a sufficient and necessary condition for a Markovian evolution is equivalent to stating that $|\Gamma(t)|$ is a monotonically decreasing function (\emph{i.e.} $\frac d {dt}|\Gamma(t)|<0$, ${\cal N}=0$); and therefore, a sufficient and necessary condition for a non-Markovian evolution is equivalent to that $|\Gamma(t)|$ is a non-monotonically decreasing function (\emph{i.e.} $\forall {\cal N}, {\cal N}>0$).

\begin{figure*}[h]
\centering
\includegraphics[width=8cm]{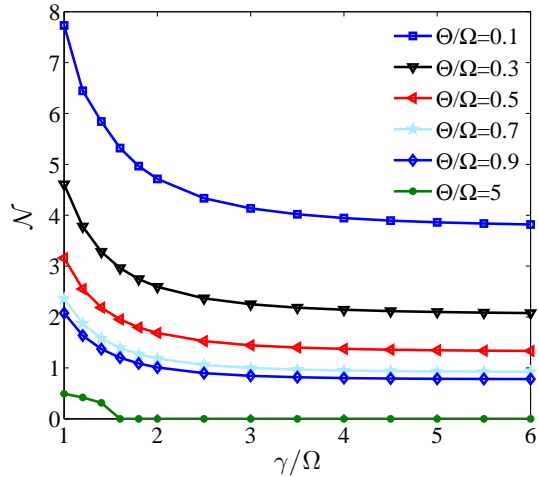}
\caption{The non-Markovianity, $\cal N$, as a function of $\gamma/\Omega$ for different values of $\Theta/\Omega$. From top to bottom, $\Theta/\Omega$ takes on values from 0.1 to 5.}
\label{fig.333}
\end{figure*}

Here we employ a pair of Pauli observables --- $\hat{\sigma}_x$ and $\hat{\sigma}_z$ --- as the incompatible measurements. These two matrices are also conventionally used to describe the spin-1/2 observables. Each of the matrices yield the eigenvalues $\pm 1$ with eigenstates $|\pm X\rangle=(|e\rangle\pm|g\rangle)/{\sqrt2}$ and $|\pm Z\rangle=\{|e\rangle,\ |g\rangle\}$. For the two Pauli operators, the uncertainty for measuring the two observables can be quantified by the entropic sum
\begin{align}
{S}_{x,z}:=S^{\rho}(\hat{\sigma}_x)+S^{\rho}(\hat{\sigma}_z).
\end{align}
To illustrate this fact, in Fig. \ref{fig.2} we vary the amount of uncertainty and the trace distance with respect to the time ($t$) for the initial state --- which was constructed with $\theta=\pi/4$ and $\phi=\pi/8$ --- for the case of
$ \Omega =\Theta =\pi\times10^6$Hz. As shown in Fig. \ref{fig.2}, the TD decreases initially and then oscillate periodically, but eventually tends to zero at the limit of long-time.
This can be interpreted as an indicator of the system becoming non-Markovian; in this case, the information stored in the atom can not only outflow but also backflow. This is to say, the information will not only be lost to the environment, but also may be recovered to some extent.
This is indicative of the capacity of the information to bi-directionally flow between the atom and the reservoir via the cavity. Eventually the entire system becomes
dynamically balanced, which drives the qubit subsystem to an asymptotic steady state.
Notably, in the non-Markovian regime, the peak values of the TD gradually become smaller with increasing time.
This reduction of the peak value for the TD implies that the backflow information is always less than the information outflow due to dissipation. To
clarify how the system evolves with fixed $\theta$, in Fig. \ref{fig.333} we plot $\cal N$ (representative of the system's non-Markovian character) as a function of $\gamma/\Omega$ for different values of $\Theta/\Omega$. From this one can
infer that there are two main factors which influence the non-Markovianity of the system: 1) the ratio value of $\gamma/\Omega$; 2) $\gamma$, which is related to the correlation time ($\tau$)
of the structured reservoir. Specifically, a stronger coupling strength, $\Omega$, between atom and cavity can lead to a greater non-Markovian character for the atomic system; \emph{a contrario}, the larger values of $\gamma$ (the longer correlation
time, $\tau$) facilitates greater non-Markovianity.

\begin{figure*}[h]
\centering
\includegraphics[width=16cm]{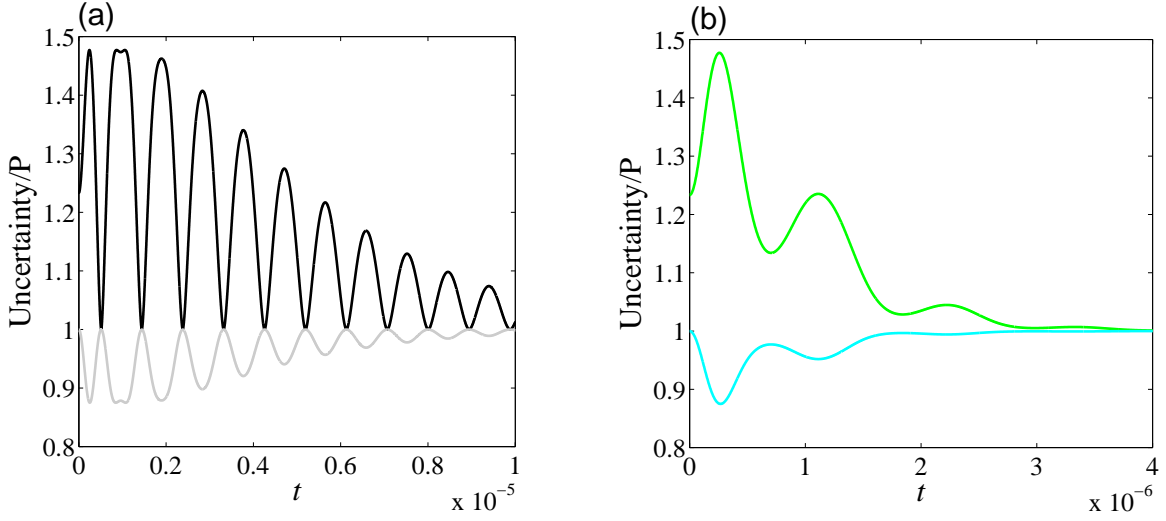}
\caption{The purity, $P$, and entropic uncertainty as a function of $t$, where we have set $\Omega=\gamma =\pi\times10^{6}$ Hz with the initial state constructed with $\theta=\pi/4$ and $\phi=\pi/8$.
In Graph (a), $ \Theta/\Omega=0.5$; the black line represents the uncertainty, while the grey line represents the purity of the atomic evolutive state. In Graph (b), $ \Theta/\Omega=5$; the green line represents the uncertainty, while the cyan line represents the purity of the atomic evolutive state.}
\label{fig.33}
\end{figure*}

Let us now shift topics to the problem of how the noise may affect the uncertainty. Intuitively, the uncertainty should become larger when the atomic subsystem moves to a mixed state from
a pure one. We plot the evolution of the measurement uncertainty with respect to time in Figs. \ref{fig.2}(a) and \ref{fig.2}(b) with $\gamma=1000\Omega$ and $\gamma=\Omega$, respectively. One can infer that: (1) In the short-time regime, the TD of the atom decreases monotonously, while the uncertainty initially increases and then decreases. Intuitively, the system will degrade when the TD decreases, and thus the uncertainty ought to constantly increase all the time, yet this disagrees with the results displayed within Figs. \ref{fig.2}(a) and \ref{fig.2}(b).
(2) The uncertainty initially increases and then shows a quasi-periodic oscillation which shrinks to the lower bound ($B_{CP}$) of the optimal uncertainty relation, and the minimal value of the uncertainty is $B_{CP}\equiv1$ as $c=\tilde{c}=\frac12$ for our choice of incompatible measurements ($\sigma_x$ and $\sigma_z$). That is to say, the uncertainty relation for the two-component measurement --- when coupled with a structured reservoir in presence of quantum memory --- never violates any previously suggested form of the uncertainty relation. This result certifies that the EUR --- as it was previously proposed --- is applicable to both the presence and absence of noises. (3) After the first minimal TD, the frequency of the uncertainty oscillation is the same as that of the TD.
This shows that the fluctuation of the uncertainty is not synchronized with the change of the atom-system TD in short-time limit, yet is synchronized with the TD after the first minimal distinguishability.
(4) The smaller $\gamma$-value can lead to the stronger non-Markovian characteristic. Stated otherwise, longer correlation times, $\tau$, of the reservoir are responsible for non-Markovianity in such a system.

To better understand the dynamics of the entropic uncertainty in the current model, we introduce the purity of a state, expressed as
\begin{align}
P={\rm Tr}(\rho^2).
\end{align}
We plot the purity and the uncertainty as a function of time in Fig. \ref{fig.33} with $\Omega=\gamma =\pi\times10^{6}$ Hz, for an initial state constructed with $\theta=\pi/4$ and $\phi=\pi/8$.
We have set $\Theta/\Omega=0.5$, and $ \Theta/\Omega=5$ in Figs. \ref{fig.33}(a) and \ref{fig.33}(b), respectively. From Figs. \ref{fig.33}(a) and \ref{fig.33}(b),
one can infer that: (1) The ratio $\Theta/\Omega$ is considerably effective at generating systemic non-Markovianity. To be
explicit, the stronger coupling strength between the atom and the cavity, $\Omega$, is responsible for non-Markovianity, while
the weaker coupling strength between the reservoir and the cavity, $\Theta$, can lead to Markovianity. This can be interpreted as
the cavity merely being another sub-environment in addition to the structural reservoir. With this in mind, one can say that both the cavity and the reservoir (which can be regarded as the total environment) can effect the non-Markovianity of the atom system.
(2) The uncertainty is fully anti-correlated with the purity of the qubit, which is a very interesting result and is consistent with previous claims in \cite{Z. Y. Xu}. This implies that
the uncertainty will increase correspondingly while the purity decreases, and vice versa.

\begin{figure*}[h]
\centering
\includegraphics[width=14cm]{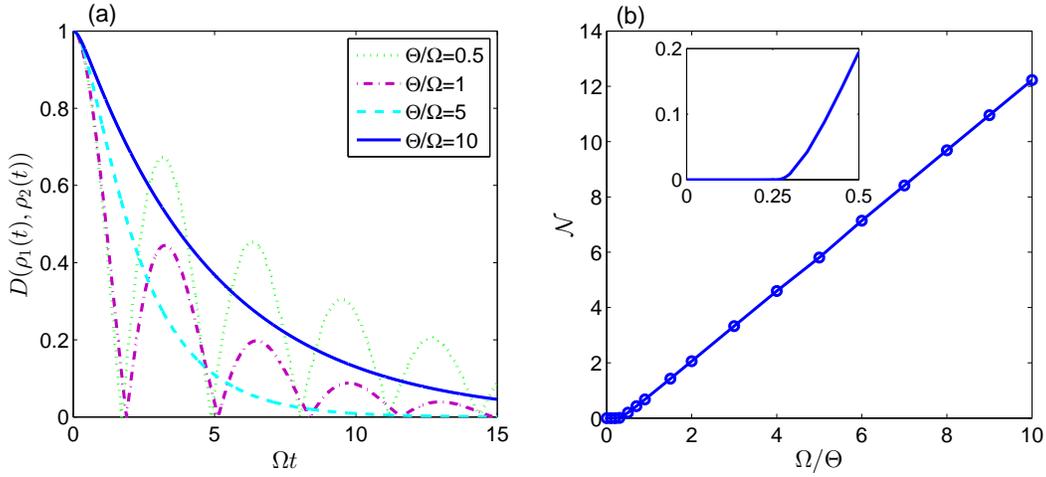} 
\caption{In Graph (a) we plot the trace distance, $D((\rho_1(t),\rho_2(t))$, with respect to the dimensionless time $\Omega t$ for various coupling-strength ratios ($\Theta/\Omega$) for an initial state constructed with $\theta=\pi/3$ and $\phi=\pi/6$. The green dotted line is plotted with $\Theta/\Omega=0.5$, the magenta dash-dotted line is for $\Theta/\Omega=1$, the cyan broken line is for $\Theta/\Omega=5$ and the blue solid line is for $\Theta/\Omega=10$. In Graph (b) we plot the non-Markovianity ($\cal N$) with respect to $\Omega/\Theta$ for an initial state constructed with $\theta=\pi/3$ and $\phi=\pi/6$. The line is broken at $\Omega/\Theta=0.25$, which is a singular point.}
\label{fig.4}
\end{figure*}

\bigskip

\noindent {\bf EUR under a memoryless reservoir.}
We shall next consider the other limiting condition: that the reservoir is memoryless, \emph{i.e.} $\tau=0\ (\gamma\rightarrow\infty)$.
In this case, the cavity's presence is solely responsible for the non-Markovian character, and the correlation time is zero. By considering $\gamma\rightarrow\infty$, one can obtain $\Gamma(t)$ in Eq. (\ref{Eq.9})
can be reduced into
\begin{align}
&\Gamma(t)=e^{-\Theta t/4} \left[\frac \Theta \lambda {\rm sinh}
\left(\frac {\lambda t}4\right)+{\rm cosh}\left(\frac {\lambda t}4\right)\right],
\end{align}
where $\lambda=\sqrt{\Theta^2-16\Omega^2}$. This expression is in agreement with the results
presented in \cite{B. Vacchini}, apart from a difference in units.
This coincidence is attributable to from the fact that the dynamics of a single qubit coupled to a vacuum reservoir with a Lorentzian spectrum could be
simulated by a pseudomode approach with a memoryless reservoir \cite{L. Mazzola,J. Jing}. Two distinct dynamical regimes are identified and undertake a
phase transition to each other at the critical condition: $\Omega_{cr}=\Theta/4$ \cite{E. M. Laine}. In the weak-coupling regime, $\Omega<\Omega_{cr}$, one can easily determine that
the dynamics are Markovian and the TD for the optimal pair ($\{|+\rangle\langle+|,|-\rangle\langle-|\}$) decreases as $\Gamma(t)$ decreases monotonically. In the strong-coupling regime, $\Omega>\Omega_{cr}$, the evolution is non-Markovian and
$\Gamma(t)$ oscillates between positive and negative values.

\begin{figure*}[h]
\centering
\includegraphics[width=14cm]{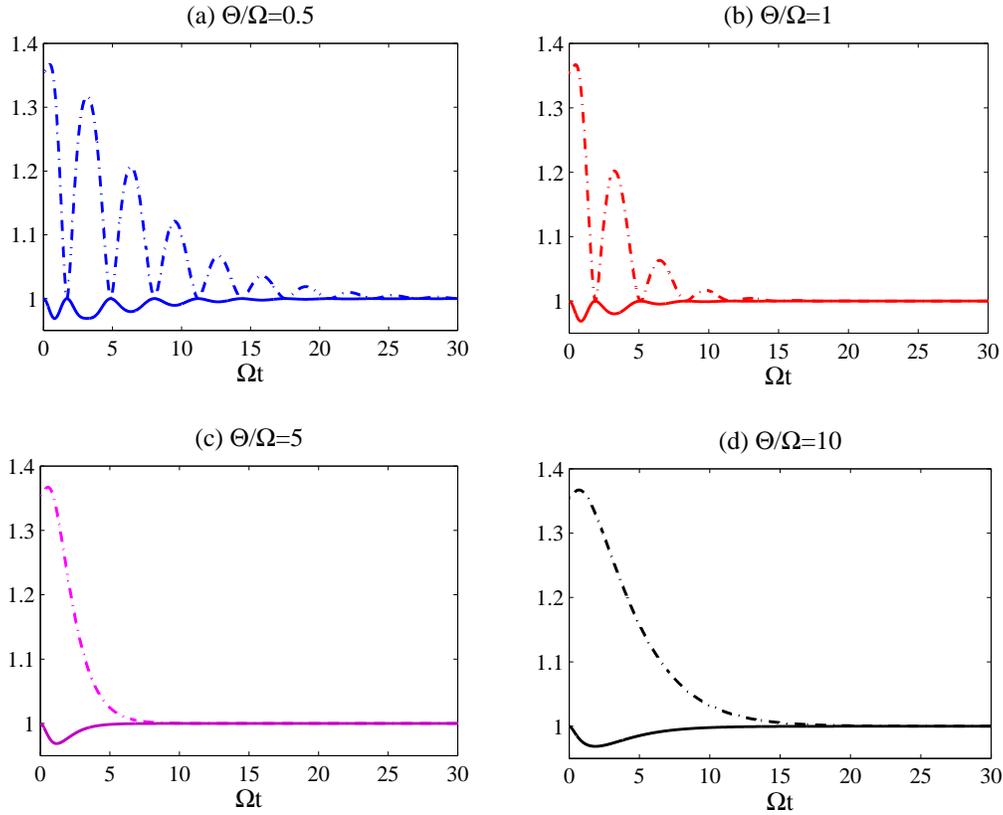}
\caption{The variation of the uncertainty for the measurement and the purity of the evolutive system --- in the absence of quantum memory --- with respect to the dimensionless time ($\Omega t$) for different coupling-strength ratio of $\Theta/\Omega$ for an initial state constructed with $\theta=\pi/3$ and $\phi=\pi/6$.
In the Figure, the dash-dotted lines represent the uncertainty, and the solid lines represent the purity of the evolutive state of the qubit. Graphs (a), (b), (c) and (d) are plotted with $\Theta/\Omega$ set to 0.5, 1.0, 5.0 and 10, respectively.}
\label{fig.55}
\end{figure*}

In what follows, we discuss how the coupling constants ($\Omega$ and $\Theta$) can influence the value of the uncertainty associated
with the measurement. As before, we employ the observable pair $\hat{\sigma}_x$ and $\hat{\sigma}_z$ as the pair of incompatibility measurements.
Let us first consider the variation of the uncertainty and the TD for the evolutive atom state
with respect to $\Omega t$. As shown in Fig. \ref{fig.4}(a), with fixed $\Theta$ the TD decreases at
first and then oscillates periodically when $\Theta/\Omega=0.5$ or 1. This can be interpreted as the
information not only flowing out of the atom, but also back-flowing into atom when $\Omega$ is sufficiently large, and hence the evolution of the atom
is non-Markovian.
A relatively small ratio of $\Omega/\Theta$ indicates that the qubit is losing information at a far slower rate than the evolution of the environment, therefore backflow of information does not occur happen
and the environment's evolution is not appreciably interrupted. When the evolution is Markovian, $\Omega<\Omega_{cr}=\Theta/4$, the dominant
effect is information outflow from the atomic system into environment, and thus TD will be reduced gradually. We plot the change of non-Markovianity
with respect to $\Omega/\Theta$ in Fig. \ref{fig.4}(b). From the Fig. \ref{fig.4}(b), the non-Markovianity ($\cal N$) is zero-valued when $\Omega/\Theta<0.25$, as the evolution of the qubit is Markovian in this situation. $\cal N$ is non-zero while $\Omega/\Theta>0.25$, implying that the evolution is non-Markovian. During a
non-Markovian evolution while $\Omega>\Omega_{cr}$, the information will not only outflow, but also backflow with increasing time.
Notably in the non-Markovian regime, the maximum value of the TD is always below unity; this limit is largely due to dissipation effects. Additionally, the entropic uncertainty increases while the TD of the atom system decreases in the short-term due to the increase in the entropic uncertainty when the system becomes unstable and undergoes dissipation. However, from Figs. \ref{fig.4} and \ref{fig.55} one can see that with the decrease of the TD, the uncertainty of measurement will firstly increase and then decrease in a relatively short-time regime. Furthermore, the magnitude of the entropic sum
undergoes periodic oscillations associated with the oscillating TD, and shrinks to the lower bound of EUR ($B_{CP}$) in the long-time regime. This indicates that the entropic uncertainty is not merely synchronous with the evolution of the atomic system at the initial stage of evolution, it becomes increasingly synchronous with the evolution of the atomic system after the TD reaches the first minimum. We note that the fluctuations of both the TD and the uncertainty become smaller as $\Theta$ grows larger, \emph{i.e.} a stronger coupling constant between the cavity and the reservoir will decrease disturbance on the entropic uncertainty. This implies that the cavity-reservoir coupling strength, $\Theta$, may dramatically influence the entropic sum. Furthermore, we plot the purity as a function of $\Omega{t}$ with different coupling-strength ratio of $\Theta/\Omega$ in Fig. \ref{fig.55} when the initial state of the qubit system is generated with $\theta=\pi/3$ and $\phi=\pi/6$. From Fig. \ref{fig.55}, it is obvious that the uncertainty is always anti-correlated with the purity of system, which is entirely consistent with our previous statement. Through the above analysis, we can conclude that stronger $\Omega$-coupling can affect the reservoir and can result in backflow of information to the atom, leading to a periodic evolution of the uncertainty.

\begin{figure*}[h]
\centering
\includegraphics[width=8cm]{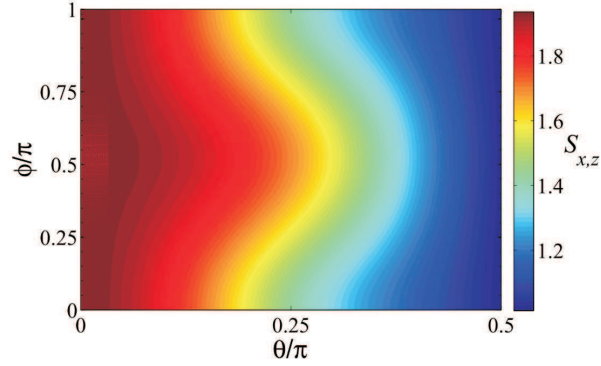}
\caption{The variation of entropic sum, $S_{x,z}$, with respect to the polar angular ($\theta$) and phase ($\phi$) of the initial state constructed with $\Omega t=10$ and $\Theta t=1.5$.}
\label{fig.6}
\end{figure*}
We also explore the relation between the initial state and the entropic sum in Fig. \ref{fig.6}, where one finds that the value of $S_{x,z}$ is symmetric about $\phi=\pi/2$, and decreases with an increase in $\theta$ for a fixed $\phi$. Specially, $S_{x,z}$ reaches a peak when $\theta=0$ and at the point of
$B_{CP}$ at $\theta=\pi/2$. This implies the excited state of the atom is more sensitively to the uncertainty of the measurement in the current model comparing with that of the ground state.

\bigskip

\noindent {\bf Reducing the uncertainty via weak measurement.}\ \ \
A novel idea has recently been proposed to protect
a state from decoherence by using quantum partially collapsing measurements, \emph{i.e.} weak measurement reversals (WMR) \cite{S. C. Wang,X. Xiao,Y. Aharonov}.
The WMR procedure is described as
\begin{align}
&\rho_{ee}(t)\rightarrow\frac{({1-m})}{{\cal C}}\rho_{ee}(t),\quad
\rho_{eg}(t)\rightarrow\frac{\sqrt{1-m}}{{ \cal C}}\rho_{eg}(t),\nonumber\\
&\rho_{ge}(t)\rightarrow\frac{\sqrt{1-m}}{{ \cal C}}\rho_{ge}(t),\quad
\rho_{gg}(t)\rightarrow\frac{1}{{ \cal C}}\rho_{gg}(t).
\end{align}
Within the above, the measurement strength $m$ satisfies $0\leq m\leq 1$ and ${\cal C}=({1-m})\rho_{ee}(t)+\rho_{gg}(t)$ is the normalized coefficient of the time-dependent state. The WMR essentially makes a post-selection that removes the result of the qubit transition $|e\rangle\rightarrow|g\rangle$; WMR can be implemented by an ideal detector to monitor the environment. This is also referred to as null-result WMR because the detector does not report any signal. In a WMR, complete collapse to an eigenstate does not occur, and thus the qubits continue in their evolution. Decoherence can be largely suppressed within the systems by uncollapsing the quantum state, returning it to the excited state.
\begin{figure*}[h]
\centering
\includegraphics[width=8cm]{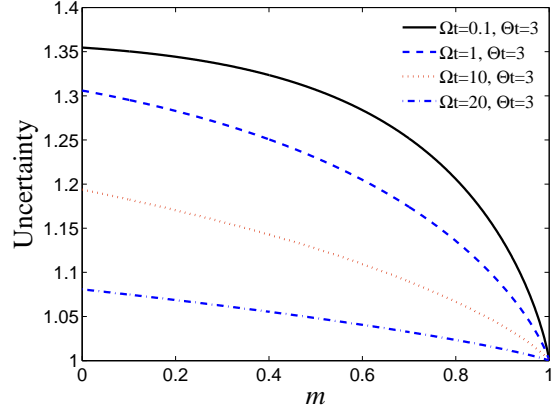}
\caption{The entropic sum, $S_{x,z}$, is plotted as a function of the measurement strength ($m$) for different coupling constant values, both $\Omega$ and $\Theta$, for a fixed real initial states constructed with $\theta=\pi/3$ and $\phi=\pi/6$.
In the Figure, the black solid line, blue broken line, red dotted line and magenta dash-dotted line represent the following values of $(\Omega, \Theta)$: (0.1, 3), (1, 3), (10, 3) and (20, 3), respectively.}
\label{fig.7}
\end{figure*}

\begin{figure*}
\centering
\includegraphics[width=14cm]{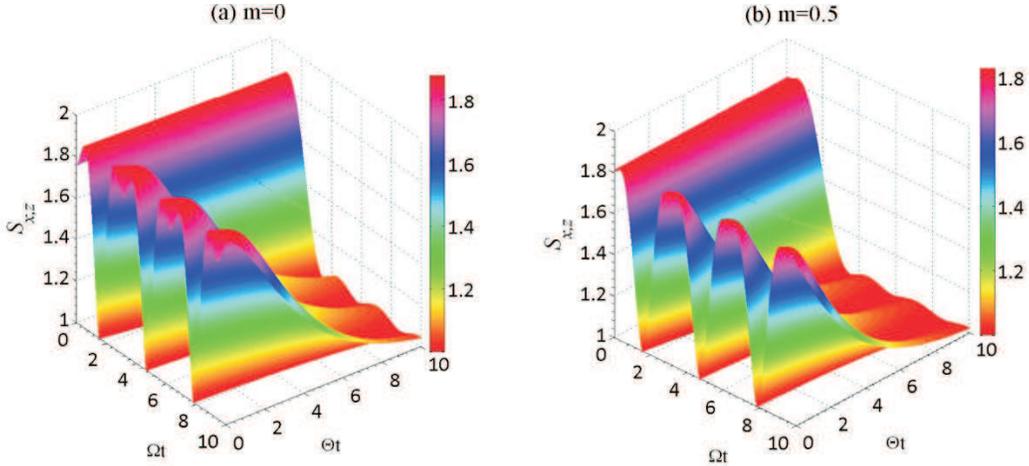}
\caption{The variation of entropic sum, $S_{x,z}$, as a function of $\Theta t$ and $\Omega t$ for different measurement strengths ($m$) with an atomic initial state constructed with $\theta=\pi/5$ and $\phi=\pi/3$.
Graph (a): $m=0$; and Graph (b): $m=0.5$.}
\label{fig.8}
\end{figure*}

It is well known that the amount of the uncertainty is crucial for quantum precision measurements, and one always expects a smaller measurement uncertainty when obtaining exact measurements. Motivated by this, we explore a methodology to reduce the uncertainty by the using appropriate WMR. For clarity, we plot the relationship between the measurement parameter $m$ and the entropic sum in Fig. \ref{fig.7}, with $\theta=\pi/3$ and $\phi=\pi/6$. From Fig. \ref{fig.7}, one can readily infer that the uncertainty decreases with the increase of the measurement strength $m$. Therefore, the WMR is capable of suppressing the decay of the atomic state, and thus largely reducing the entropic uncertainty during the crossover from Markovianity to non-Markovianity. Furthermore, we investigate the relation between the entropic uncertainty and the coupling strengths $\Theta$ and $\Omega$ in Fig. \ref{fig.8} for
$\theta=\pi/5$ and $\phi=\pi/3$, both with and without weak measurement ($m=0.5$). It is obvious that the maximal value of the uncertainty in the case $m=0.5$ is smaller than that of $m=0$, which indicates that WMR can efficiently reduce the uncertainty of measuring a pair of incompatible observables. Furthermore, Figs. \ref{fig.8}(a) and (b) show that the uncertainty will vary periodically with respect to the coupling strength $\Omega t$, consistent with the previously obtained results.

{\large \noindent {\bf Conclusion}}\\
Herein, we investigate how a bosonic environment influences the uncertainty of measuring two incompatible measurements on an atom-cavity coupled system during the crossover between Markovianity and non-Markovianity. Notably, in the presence of memory effects the evolution of the atom system is determined by the strength of the cavity and the structured reservoir. The uncertainty is characterized by fluctuations which are not synchronized with the change of the systemic state, tending to the lower bound in the long-time limit. In the absence of memory effects, we numerically verified that the amount of EUR is correlated with the coupling strengths of the atom-cavity and the cavity-reservoir. We find that the coupling strengths of the atom-cavity and the cavity-reservoir greatly influences the uncertainty and its dynamic behavior. The relatively strong coupling strength between the cavity and the structured
reservoir can provide a natural reduction of the overall uncertainty. Additionally, we conclude that the stronger atom-cavity coupling strength results in information backflow to the atom manifesting itself as an oscillation in the uncertainty. Explicitly, the uncertainty oscillates to the lower bound of EUR when $\Omega>\Omega_{cr}$; the uncertainty will reduce all the time and shrink to the lower bound in the long-time regime when $\Omega<\Omega_{cr}$.
We have also verified that the uncertainty for the measurement is anti-correlated with the purity of the evolutive qubit state, whether the system is Markovian or non-Markovian. Notably, we propose an efficient method to reduce the uncertainty for a pair of observables with such system via post-selection weak measurement reversal. Therefore, our investigation may shed light on the generation of precision measurements for a system coupled with a multi-degree-of-freedom environment possessing either Markovian
or non-Markovian character.

\bigskip

{\large \noindent {\bf Methods}}\\
Here, we deal with the reduced dynamics of the atomic subsystem. Assuming that both the cavity and environmental reservoir are initially in their vacuum states.
The model can be solved analytically and thus can fully capture the features of the atomic subsystem.
In the one-excitation subspace, the total state can generally be written as \cite{H. P. Breuer}
\begin{align}
|\Psi(t)\rangle=a(t)|g,0,0_j\rangle+b(t)|e,0,0_j\rangle+c(t)|g,1,0_j\rangle
+\sum_jh_j(t)|g,0,1_j\rangle,
\label{Eq.66}
\end{align}
where $|0\rangle$ and $|1\rangle$ are the vacuum and single-photon states of the cavity, while $|0_j\rangle$  and $|1_j\rangle$the cavity represent no excitation and one excitation in the $j$th mode of the reservoir. In what follows, we derive the coefficients of the state of the composite system. Substituting Eq. (\ref{Eq.66}) into the Schr\"{o}dinger equation
\begin{align}
i\frac{d}{dt}|\Psi(t)\rangle=\mathcal{H}_I|\Psi(t)\rangle,
\label{Eq.6}
\end{align}
yields the following formulae
\begin{eqnarray}
{a(t)}&=&a(0),\quad  \frac{d}{dt}b(t)=-i\Omega c(t), \quad \frac{d}{dt}h(t)=-i\Delta_j e^{i\delta_jt}c(t),\notag\\
\frac{d}{dt}c(t)&=&-i\Omega b(t)-i\sum_k\Delta_je^{-i\delta_jt}h_j(t)d\tau.
\end{eqnarray}
Linking the initial conditions $c(0)=h_j(0)=0$ with the correlation function $\alpha(t,s)=\sum|\Delta_j|^2e^{-i\delta_j(t-s)}=\frac{\Theta\gamma}{2}e^{-\gamma|t-s|}$, one can exactly obtain the atomic dynamics by means of tracing out both the cavity and the reservoir subsystem, \emph{i.e.} $\rho={\rm Tr}_{C,R}[|\Psi(t)\rangle\langle\Psi(t)|]$. In this way, one can derive the desired reduced matrix of the atomic state, as is in Eq. (\ref{eq.7}).

\vspace{.5cm}

{\large {\noindent {\bf Acknowledgements}}} \\
This work was supported by the National Natural Science Foundation of China (Grant Nos. 61601002, 61275119, and 11575001), Anhui Provincial Natural Science Foundation (Grant No. 1508085QF139), the fund of National Laboratory for Infrared Physics (Grant No. M201307), and is a project from National Science Foundation Centers for Chemical Innovation: CHE-1037992.

%
%

\vspace{.5cm}

{\large {\noindent {\bf Additional Information }}} \\
{\bf Competing financial interests:} The authors declare no competing financial interests.


\end{document}